# INHOMOGENEITY OF MAGNESIUM DIBORIDE STRUCTURE AND ITS EFFECT ON CRITICAL CURRENT DENSITY

Prikhna T., Gawalek W.*, Savchuk Ya., Kozyrev A., Wendt M.*, Dellith J.*, Goldacker W.[†], Dub S, Sergienko N., Habisreuther T.*, Moshchil V., Dittrich U.[‡], Karau W.[‡], Noudem J.[††], Schmidt Ch.*, Turkevich V., Litzkendorf D.*, Weber H.**, Eisterer, M.**, Melnikov V., Nagorny P., Sverdun V.

Institute for Superhard Materials of the National Academy of Sciences of Ukraine, 2, Avtozavodskaya Str., Kiev 07074, Ukraine
* Institut für Photonische Technologien, Albert-Einstein-Strasse 9, Jena, D-07745,Germany
[†] Forschungszentrum Karlsruhe, Institut für Technische Physik, Postfach 3640, D-76021 Karlsruhe
[‡] H.C. Starck GmbH, Goslar 38642, Germany
[††] CNRS/CRISMAT/ISMRA, 6, Bd du Maréchal Juin, CNRS UMR 6508, 14050 Caen, France
** Atomic Institute of the Austrian Universities, 1020 Vienna, Austria

The influence of the oxygen segregation into Mg-B-O inclusions (which are oxygen enriched as compared to the matrix material) and the formation of boron enriched phases (with near $MgB_{12}$ composition) in $MgB_2$ structure on pinning in magnesium diboride and the effect of SiC and Ti alloying on the formation of structural inhomogenities in $MgB_2$ are discussed. A nanostructural material of near theoretical density (2.7 g/cm$^3$) based on $MgB_2$-SiC (10 wt% of 200-800 nm SiC was added), which is characterized by high $j_c$ ($10^6$ A/cm$^2$ in 1 T - $10^5$ A/cm$^2$ in 3.5 T at 20 K and $10^5$ A/cm$^2$ at 35 K) has been synthesized at 2 GPa.

INTRODUCTION

It is well known that alloying of $MgB_2$ with SiC, Ti, and Zr can increase the critical current density, $j_c$, of the material after synthesis under ambient pressure [1, 2]. It has been shown by us that the alloying with Ti, Ta and Zr can increase $j_c$ of the high pressure (HP) synthesized or sintered at 2 GPa $MgB_2$-based material [3-5]. HP conditions allow us not only to suppress oxygen evaporation and promote the material densification thus improving connectivity between grains, but to change the kinetics and thermodynamics of the process as well. In the case of HP synthesis in the presence of additions, the $j_c$ increase due to the absorption by Ti, Ta and Zr of admixing hydrogen (see, for example Figure 1 e, f), which prevents the formation of $MgH_2$ impairing the SC characteristics, and due to the fact that these additions stimulate the increase of the volume of boron-enriched Mg-B inclusions (with near $MgB_{12}$ stoichiometry), which are relevant pinning centers in $MgB_2$ [6, 7]. It seems that the presence of these metals may reduce the nucleation energy of the boron-enriched phases ($MgB_{12}$, for example) and their grains may form in the regions of high boron concentration.

Up to now our attempts to improve $j_c$ of HP-synthesized $MgB_2$ by adding SiC with 20-30 and 200-400 nm grain sizes were unsuccessful: $j_c$ at 10 - 35 K decreased or in the best cases remained unchanged [5]. But when HP-synthesized $MgB_2$ was doped with SiC with 200-800 nm grains an essential increase of $j_c$ up to the record values occurred and in this case the $Mg_2Si$ phase was not observed in the X-ray pattern (Fig. 1 d). Besides, here we report our new observations concerning the formation of Mg-B-O inclusions of size about 50-200 nm that are oxygen-enriched as compared to the oxygen content of the matrix. The possible influence of oxygen-enriched Mg-B-O inclusions on pinning in the $MgB_2$ as well as the effect of SiC and Ti alloying on the oxygen segregation in the material are discussed.

EXPERIMENTAL

In the experiments on synthesis, metallic Mg chips (technical specifications 48-10-93-88) and amorphous B (H.C. Starck, Type 1 contains 0.66% oxygen, and Type 2 contains 1.5% oxygen with grains of 4 μm and <5 μm, respectively, the amount of crystalline phase was higher in Type 1) have been taken in the stoichiometric ratio of $MgB_2$. To study the influence of Ti and SiC, a powder of Ti (Ti- 30 μm, 95 % purity) or SiC (with particle sizes of 200-800 nm, H.C.Starck) has been added to the stoichiometric mixture of Mg and B in amounts of 10 wt%. Then we mixed and milled the components in a high-speed activator with steel balls for 1-3 min. The resulting powder was compacted into tablets. The X-ray study of the initial Mg, Ti and B showed that the materials contained no impurity phases with hydrogen (the method accuracy being about 3-5%). The high pressure (2 GPa) - high temperature (700-1100 °C) conditions for 1 h were created in the recessed-anvil type high-pressure apparatuses (HPA) described elsewhere [8]. During the synthesis samples were in contact with a precompacted powder of hexagonal BN. The structure of the materials was studied using X-ray diffraction and a SEM equipped with microprobe analyzer. All $j_c$ data were estimated from magnetization hysteresis loops obtained with an Oxford Instruments 3001 vibrating sample magnetometer (VSM) using Bean's model. For VSM measurements, samples with a typical diameter of 3 mm were prepared and this size was applied to calculate the $j_c$ -values.

RESULTS AND DISCUSSIONS

The nanostructural material (the average grain size being 15-20 nm) of near theoretical density (2.7 g/cm$^3$) based on $MgB_2$-10 wt % of SiC, which is characterized by the record high $j_c$ has been HP synthesized using Type 1 boron (see Figure 1a). The critical current density, $j_c$, in zero magnetic field is $1.6 \times 10^6$ A/cm$^2$ at 10 K and $1.24 \times 10^6$ A/cm$^2$ at 20 K $j_c$ remained above $10^6$ A/cm$^2$ in the fields up to 2 and 1 T and higher than $10^5$ A/cm$^2$ in the fields up to 5 and 3.5 T at 10 and 20 K, respectively. At 20 K the field of irreversibility exceeds 8 T, and at 10 K in 10 T field $j_c$ is higher than 1000 A/cm$^2$. At 35 K in zero magnetic field $j_c$ was about $10^5$ A/cm$^2$ and the field of irreversibility was above 1 T. According to X-ray data (Fig. 1b), the material contains $MgB_2$, SiC and a small amount of MgO. $MgB_2$ reflexes are sharp

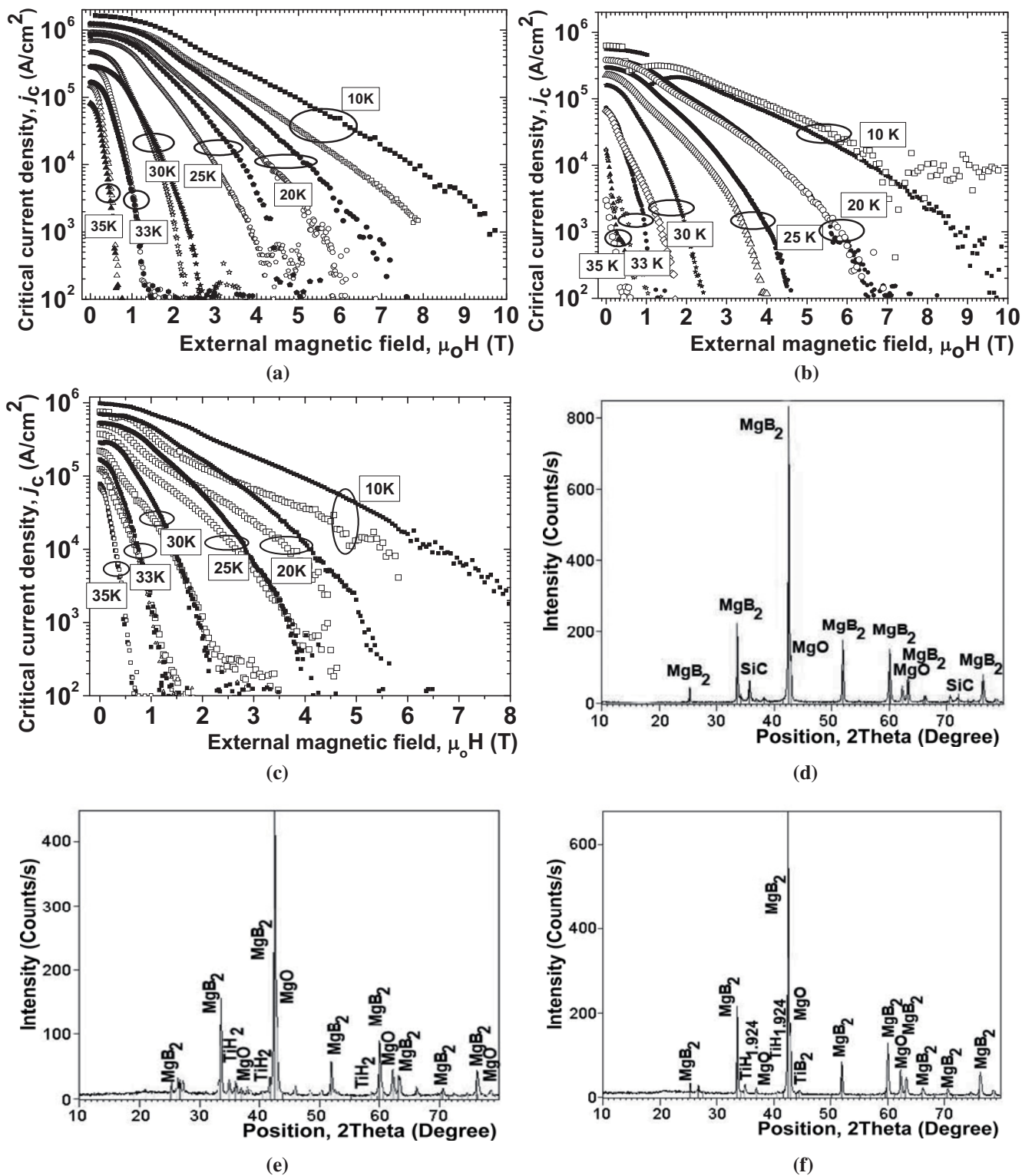

Figure 1 (a)–(c) Dependences of critical current densities, $j_c$, on the external magnetic fields, $\mu_o H$, at temperatures 10, 20, 25, 30, 33 and 35 K for the samples synthesized (a) at 1050 °C, 2 GPa for 1 h from B (Type 1) and Mg without additions (opened symbols) and with 10 % of SiC (solid symbols); (b) at 2 GPa for 1 h from B (Type 2) and Mg with 10 % of SiC at 1050 °C (solid symbols) at 800 °C (opened symbols), (c) at 2 GPa for 1 h from B (Type 2) and Mg with 10 % of Ti at 1050 °C (solid squares) and at 800 °C (opened squares); (d)–(f) X-ray patterns of the sample synthesized (d) at 1050 °C, 2 GPa for 1 h from B (Type 1) and Mg with 10 % of SiC; (e) from B (Type 2) and Mg with 10 % of Ti at 2 GPa for 1 h at 800 °C and (f) at 1050 °C.

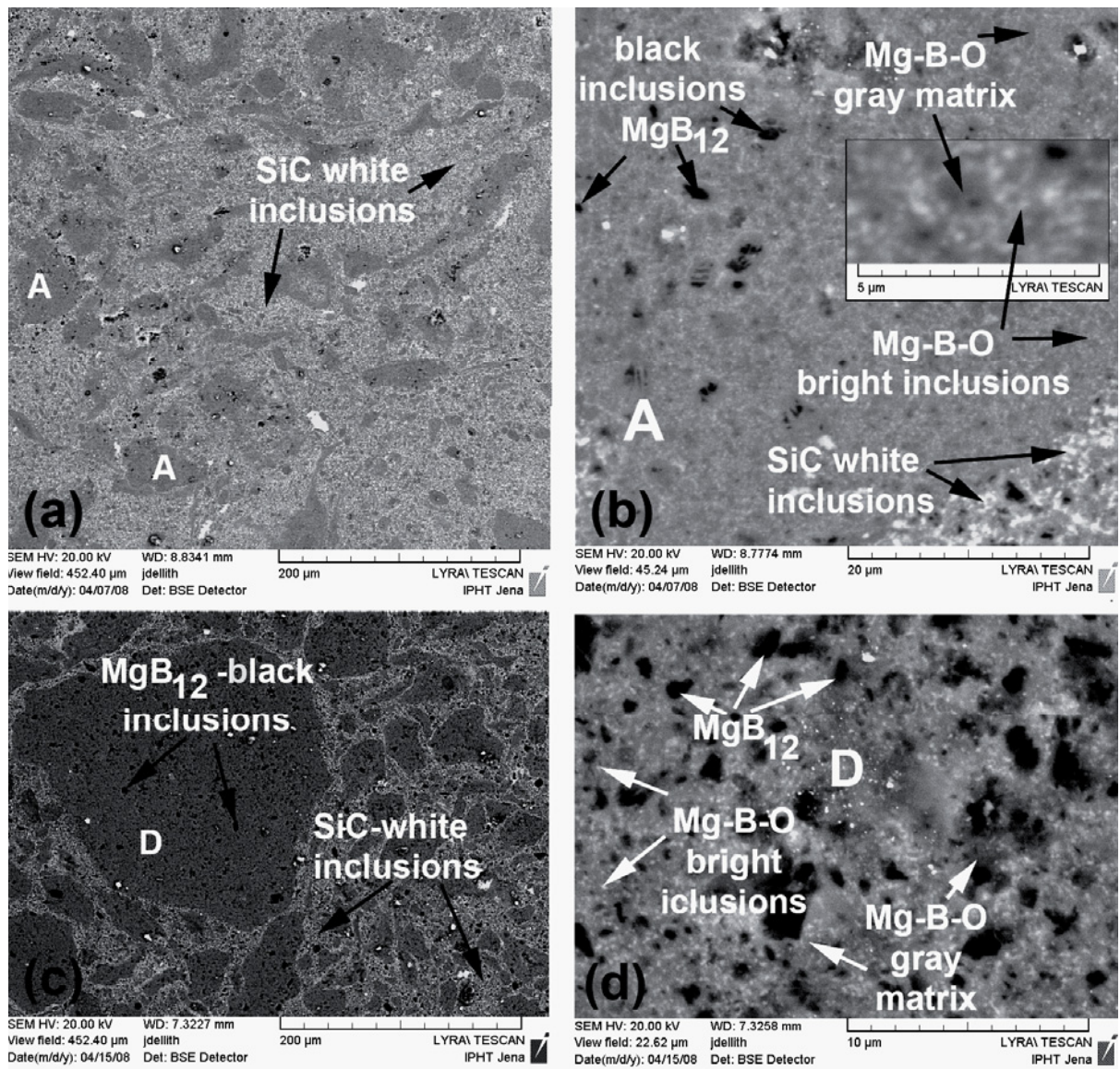

Figure 2 Structure obtained by SEM (COMPO –backscattering electron image) of the sample synthesized at 1050 °C, 2 GPa during 1 h from boron Type 1 and magnesium with SiC at (a) 500 power and (b) 5000 power magnifications (area marked as "A" in Fig. 2a). The inset in Fig. 2b shows the same area "A" at 20000 power magnification. Fig 2d shows site "D" where SiC inclusions are absent (at 10000 power magnification).

and are not shifted from the reference positions, there are no $Mg_2Si$ reflexes, as it was in the case of SiC with 20-30 or 200-400 nm grains [5]. The structure of materials with 200-800 nm SiC added manufactured from both types of initial boron (Figure 2) contained boron-enriched inclusions, whose stoichiometry was near to $MgB_{12}$. It is known that $MgB_{12}$ reflexes are absent in the X-ray pattern due to poor diffracted signals because of the low X-ray atomic scattering factor of boron [9].

It seems that the absence of the interaction and comparatively homogeneous distribution of SiC when Type 1 boron was used (Figure 2a) can be one of the reasons for the high $j_c$ of the material. A somewhat lower $j_c$ was exhibited material produced from Type 2 boron with less homogeneously distributed SiC (Fig. 2b) grains. The SC properties of the material from Type 2 boron with SiC addition were essentially improved and rather high absolute $j_c$ values were obtained (Fig. 1 b). It should be mentioned that this material showed higher $j_c$ at low temperatures after synthesis at 800 °C (opened symbols in Figure1b) and at higher temperatures after synthesis at 1050 °C (solid symbols on Figure 1b).

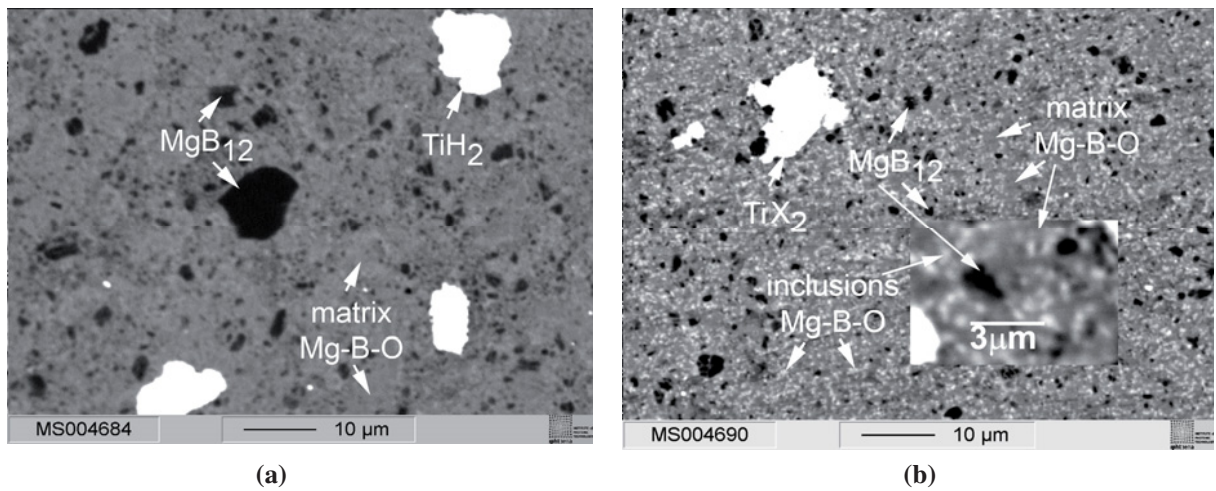

Figure 3 Structure of the samples synthesized under 2 GPa for 1h at 800 °C (a) and 1050 °C (b) from boron (2B) and magnesium with 10 % of Ti added. TiX$_2$ – can be TiH$_2$ or TiB$_2$ (the only Ti was found by SEM, but according to the X-ray patterns two phases TiH$_2$ and TiB$_2$ are present in the material, see Figure 1f). The inset in Fig. 3b shows the structure of the same sample at higher magnification.

So, two optimal synthesis temperatures 800 and 1050 °C have been found.

Adding of Ti to the material prepared from the Type 2 boron we can induce much higher improvements in $j_c$ (Fig. 1c) than by SiC adding if the synthesis temperature is 1050 °C (Fig. 2b, solid symbols). The $j_c$ of the material (from Type 2 boron) synthesized at 800 °C with Ti was somewhat higher than in the case of SiC addition (Figs. 1c and 1b, opened symbols), but absolute values were much lower than that of the material synthesized at 1050 °C with Ti (Fig. 1c). However, the Ti adding to the material produced from the Type 1 boron induced the $j_c$ reduction (as compared to the $j_c$ of the material without additions) for all synthesis temperatures. Ti-containing inclusions cannot be pinning centres because they are rather big and not so homogeneously distributed (Fig. 3 a, b). It is interesting that the material produced from the Type 1 boron without additions had very high $j_c$ too, even somewhat higher than that produced from Type 2 boron with Ti at 1050 K (compare opened symbols in Figure 1a and solid symbols in Figure1c). Type 1 boron has initially lower oxygen content and some less amount of amorphous phases as compared to the Type 2 boron, but as it was shown earlier [10] no correlations was found between the amount of oxygen in the initial boron (which ranged from 0.66 to 3.5 wt.%) and its amount in the HP-synthesized or sintered materials as well as the correlations with the material $j_c$.

Interesting is that in all materials synthesized at 1050 °C the segregation of oxygen has been observed, which gives rise to Mg-B-O inclusions rather homogeneously distributed in the structure (bright inclusions in Figures 2b, d and 3b.). It was established that the oxygen content in these inclusions is higher than that in the matrix. At low synthesis temperatures the segregation of oxygen is not clearly pronounced and the matrix looks rather homogeneous (Fig. 3a). The amount of oxygen in the gray matrix of the material prepared from Type 2 boron with Ti after synthesis at 800 °C was about 8% and after synthesis at 1050 °C only 5%. To determine the oxygen content of Mg-B-O inclusions was not possible due to the limited SEM resolution, but it was higher than that in the matrix. So, as the synthesis temperature increased, the amount of higher borides usually decreased, but segregation of oxygen became more pronounced. It was found that the formation of both boron-enriched phases and Mg-B-O inclusions (due to the oxygen segregation) can be influenced by the additions of SiC and Ti.

All these observations let us conclude that structural inhomogeneities such as boron-enriched phases ($MgB_{12}$, in particular) and Mg-B-O inclusions are very important for high critical currents in $MgB_2$. Mg(O,H) inclusions that can be pining centres in smaller amount than in our case were found in $MgB_2$ structure by Liao, X. Z et al. [11]. We should not exclude the positive effect on pinning the SiC grains by themselves (when it is no notable interaction between SiC and $MgB_2$, because when $Mg_2Si$ forms carbon can go to $MgB_2$ structure [9], thus reducing $T_c$). But many experiments with different types of initial boron and magnesium diboride have assured us that such structural inhomogenities as boron-enriched phases ($MgB_{12}$ at 2 GPa) as well as Mg-B-O inclusions largely influence pinning and $j_c$ of $MgB_2$.

Besides, our specially directed experiments point to the probability of superconductive behavior of the higher borides ($MgB_{12}$, in particular). When specially synthesized samples contained mainly the $MgB_{12}$ phase (which is harder than sapphire and two times harder than $MgB_2$) they exhibited rather high SC characteristics (the detailed results will be published later). May be if $MgB_{12}$ is superconductive it also can be the reason why the $j_c$ improved when the amount of $MgB_{12}$ dispersed grains increased in $MgB_2$-based sample structure.